\documentclass[12pt]{iopart}
\usepackage{iopams}

\newcommand{\eqd}{\stackrel{\rm def}{=}}%
\newcommand{\norm}[1]{\mathop{\rm vrai\:sup}\limits_{\{q_1,\ldots,\,q_s\}
\in V} \left|{#1}\right|}%

\begin{document}

\title[On completeness of description of an equilibrium canonical
ensemble]{On completeness of description of an equilibrium
canonical ensemble by reduced s-particle distribution function}

\author{M I Kalinin}

\address{The Russian Research Institute for Metrological
Service, 46 Ozernaya str., Moscow, Russia, 119361}%
\ead{kalinin@vniims.ru}

\begin{abstract}
In this article it is shown that in a classical equilibrium
canonical ensemble of molecules with $s$-body interaction full
Gibbs distribution can be uniquely expressed in terms of a reduced
s-particle distribution function. This means that whenever a
number of particles $N$ and a volume $V$ are fixed the reduced
$s$-particle distribution function contains as much information
about the equilibrium system as the whole canonical Gibbs
distribution. The latter is represented as an absolutely
convergent power series relative to the reduced $s$-particle
distribution function. As an example a linear term of this
expansion is calculated. It is also shown that reduced
distribution functions of order less than $s$ don't possess such
property and, to all appearance, contain not all information about
the system under consideration.
\end{abstract}


%Keywords: \quad Rigorous results in statistical mechanics

%Uncomment for PACS numbers title message
%\pacs{00.00, 20.00, 42.10}
% Keywords required only for MST, PB, PMB, PM, JOA, JOB?
%\vspace{2pc}
%\noindent{\it Keywords}: Article preparation, IOP journals
% Uncomment for Submitted to journal title message
%\submitto{\JPA}
% Comment out if separate title page not required
\maketitle

\section{Introduction}

In classical statistical mechanics an equilibrium system of $N$
molecules in a volume $V$ is described by canonical distribution
function $F_N(q, p)$, where $(q, p)$ is a set of phase variables:
coordinates $q_i$ and momenta $p_i$ of molecules. If interaction
of molecules is additive, reduced distribution functions are
introduced \cite{Balescu,NN}. They are used for evaluation of
thermodynamic characteristics of such molecular system. It's
usually accepted that reduced distribution functions contain
information about a molecular system less than the canonical Gibbs
distribution function. It's also supposed that the lower an order
of a reduced distribution function is, the less information it
contains. But there does not exist a proof of this statement in
scientific literature.

On the other hand, it is known that for an equilibrium canonical
ensemble of non-interacting particles  a canonical distribution
function $F_N(q,p)$ is decomposed into a product of reduced
one-particle distribution functions $F_1(q,p)$ \cite{Balescu}.
This means that all information about such system is contained in
the reduced one-particle distribution function. In
\cite{Kalinin1,Kalinin2} it was proved that for a closed molecular
system with pair interaction there is a one-to-one correspondence
between a canonical distribution and a reduced two-particle
distribution function.

In this paper it's proved that for a system having interaction
potentials up to order $s$ there exists one-to-one correspondence
between full canonical distribution function and a reduced
$s$-particle distribution function. This means that the
$s$-particle function contains the whole information about system
under consideration.

We consider an equilibrium system of $N$ particles contained in
the volume $V$ under the temperature $T$. Potential energy of
system is supposed to have the form
\begin{equation}
\label{U_N} %
U_N(q_1,\ldots,q_N)=\sum_{l=1}^s\: \sum_{1\le j_1<\cdots<j_l\le
N}u_l(q_{j_1},\ldots,q_{j_l}),
\end{equation}                                            %  {U_N} (1)
where $s$ is an arbitrary fixed integer less than $N$ and
$u_l(q_1,\ldots,q_l)$ is a direct interaction energy of $l$
particles. Probability distribution function of equilibrium system
is the canonical Gibbs distribution which is decomposed into a
product of a momentum distribution function and a configurational
one \cite{Balescu,NN}. The former is expressed as a product of
one-particle Maxwell distributions, the latter has the form
\begin{equation}
\label{D_N} %
D_N(q_1,\ldots,q_N)=Q_N^{-1}\exp\{-\beta U_N(q_1,\ldots,q_N)\},
\end{equation}                                            %  {D_N} (2)
where $\beta=1/kT$, $k$ is the Boltzmann constant and $Q_N$ is the
configuration integral
\begin{equation}
\label{Q_N} %
Q_N=\int\!\!\exp\{-\beta U_N(q_1,\ldots,q_N)\}dq_1 \cdots dq_N.
\end{equation}                                            %  {Q_N} (3)
Here and below integrating with respect to every configurational
variable is carried out over the volume $V$. For a system having
interaction of form (\ref{U_N}) reduced distribution functions are
introduced by expressions \cite{Balescu}
\begin{equation}
\label{F_l} %
F_l(q_1,\ldots,q_l)=\frac{N!}{(N-l)!}\int D_N(q_1,\ldots,q_N)
dq_{l+1}\cdots dq_N, \quad l=1,2,\ldots .
\end{equation}                                            %  {F_l} (4)
These functions are used instead of full canonical distribution
(\ref{D_N}) to calculate various characteristics of the molecular
system. Let us investigate properties of the reduced $s$-particle
distribution function.

Potential energy (\ref{U_N}) can be written as
\begin{equation}
\label{U_N1} %
U_N(q_1,\ldots,q_N)=\sum_{1\le j_1<\cdots<j_s\le
N}\phi(q_{j_1},\ldots,q_{j_s}),
\end{equation}                                            %  {U_N1} (5)
where
\begin{equation}
\label{phi_s} %
\phi(q_1,\ldots,q_s)=\sum_{r=1}^s (C_{N-r}^{s-r})^{-1}\sum_{1\le
j_1<\cdots<j_r\le s}u_r(q_{j_1},\ldots,q_{j_{\,r}})
\end{equation}                                            %  {phi_s} (6)
and $C_m^{\,n}=m!\bigl/\{n!(m-n)!\}$ are binomial coefficients.

Introduce a function $h(q_1,\ldots,q_s)\}$ by the relation
\begin{equation}
\label{h} %
\exp\{-\beta \phi(q_1,\ldots,q_s)\}=
\sigma\{1+h(q_1,\ldots,q_s)\},
\end{equation}                                            %  {h} (7)
where
\begin{equation}
\label{sigma} %
\sigma= \frac{1}{V^s}\int\exp\{-\beta \phi(q_1,\ldots,q_s)\}\,dq_1
\cdots dq_s.
\end{equation}                                            %  {sigma} (8)
The canonical Gibbs distribution (\ref{D_N}) takes the form
\begin{equation}
\label{D_N1} %
D_N(q_1,\ldots,q_N)=Q_N^{-1}\prod_{1\le j_1<\cdots<j_s\le
N}[1+h(q_{j_1},\ldots,q_{j_s})]
\end{equation}                                            %  {D_N1} (9)
with
\begin{equation}
\label{Q_N1} %
Q_N=\int\!\!\prod_{1\le j_1<\cdots<j_s\le
N}[1+h(q_{j_1},\ldots,q_{j_s})]\,dq_1 \cdots dq_N.
\end{equation}                                            %  {Q_N1} (10)

From (\ref{D_N1}) and (\ref{Q_N1}) it follows that statistical
properties of system under consideration are completely determined
by the specifying single function of $s$ configurational variables
$h(q_1,\ldots,q_s)$ and two external parameters $N$ and $V$. The
assumption naturally arises that another single function of $s$
configurational variables $F_s(q_1,\ldots,q_s)$ can completely
determine all statistical properties of this system. It turns out
that this is indeed the case and reduced distribution function of
order $s$ plays a special role among all reduced functions
(\ref{F_l}). It can be proved that canonical Gibbs distribution
(\ref{D_N}) is expressed in terms of $F_s(q_1,\ldots,q_s)$. This
means that there is one-to-one correspondence between $D_N$ and
$F_s$. Therefore the system under consideration can be completely
described by both canonical distribution and reduced $s$-particle
distribution function. All functions (\ref{F_l}) are expressed in
terms of $F_s(q_1,\ldots,q_s)$ too. We prove these statements
below. That proof is analogous to one for system with two-body
interaction ($s=2$) presented in the papers
\cite{Kalinin1,Kalinin2}.

In section \ref{problem} we pose a mathematical problem for our
molecular system  and formulate conditions for existence and
uniqueness of its solution. In section \ref{proof} feasibility of
these conditions for considered physical system are proved. In
section \ref{sec-h(f)} an expression for function
$h(q_1,\ldots,q_s)$ in terms of $F_s(q_1,\ldots,q_s)$ is
calculated. In section \ref{sec-phi(f)} an expression for the
canonical distribution in terms of reduced $s$-particle
distribution function is produced. In section \ref{inadequacy} it
is shown that reduced distribution functions of orders less then
$s$ don't possess such property.

\section{Mathematical formulation of problem\label{problem}}

Let us introduce a function $f(q_1,\ldots,q_s)$ by the relation
\begin{equation}
\label{f} %
F_s(q_1,\ldots,q_s)= \frac{N!}{(N-s)!\,V^s}[1+f(q_1,\ldots,q_s)].
\end{equation}                                            %  {f} (11)
 Both the function $f(q_1,\ldots,q_s)$ and the function $h(q_1,\ldots,q_s)$
 satisfy the conditions
\begin{equation}
\label{cond_f-h} %
\int f(q_1,\ldots,q_s)\,dq_1 \cdots dq_s=0, \qquad \int
h(q_1,\ldots,q_s)\,dq_1 \cdots dq_s=0.
\end{equation}                                       %  {cond_f-h} (12)

From expressions (\ref{F_l}), (\ref{D_N1}), and (\ref{f}) it
follows that
\begin{equation}
\label{h-f} %
1+f(q_1,\ldots,q_s)=\frac{V^s}{Q_N}\int\prod_{1\le
j_1<\cdots<j_s\le N}[1+h(q_{j_1},\ldots,q_{j_s})]dq_{s+1}\cdots
dq_N.
\end{equation}                                            %  {h-f} (13)
This relation defines the transformation $\{h\rightarrow f\}$ and
can be considered as a nonlinear equation relative to
$h(q_1,\ldots,q_s)$. If there exists a solution
$h(q_1,\ldots,q_s;[f])$ of this equation then the function
$D_N(q_1,\ldots,q_N)$ becomes an operator function of $f$. It
means that both the canonical Gibbs distribution $D_N$ and all
reduced distribution functions $F_l$ are expressed in terms of the
single reduced distribution function $F_s$. Thus we have to prove
that equation (\ref{h-f}) has a unique solution and therefore the
transformation $\{h\rightarrow f\}$  has inverse one
$\{f\rightarrow h\}$.

Multiplying equation (\ref{h-f}) by $Q_NV^{-N}$ and using
(\ref{Q_N1}) we rewrite it in the form
\begin{eqnarray}
\fl [1+f(q_1,\ldots,q_s)]\frac{1}{V^N}\int\!\!\prod_{1\le
j_1<\cdots<j_s\le N}[1+h(p_{j_1},\ldots,p_{j_s})]\,dp_1 \cdots
dp_N\nonumber\\
-\frac{1}{V^{N-s}}\int\prod_{1\le j_1<\cdots<j_s\le
N}[1+h(q_{j_1},\ldots,q_{j_s})]dq_{s+1}\cdots dq_N=0.
\label{h-f_1} %
\end{eqnarray}                                            %  {h-f_1} (14)
The left-hand side of (\ref{h-f_1}) is a polynomial operator of
degree ${\cal N}=C_N^s$ relative to $h$ and of degree one relative
to $f$. Denote this operator by ${\cal F}(h,f)$. Equation
(\ref{h-f_1}) can be written in a symbolic form
\begin{equation}
\label{F(h-f_2)} %
{\cal F}(h,f)=0.
\end{equation}                                            %  {F(h-f_2)} (15)
To solve this equation it is necessary to specify an additional
condition
\begin{equation}
\label{F(h0,f0)} %
{\cal F}(h^{(0)},f^{(0)})=0,
\end{equation}                                            %  {F(h0,f0)} (16)
where $f^{(0)}(q_1,\ldots,q_s)$ and $h^{(0)}(q_1,\ldots,q_s)$ are
assigned functions.

We can easily determine these functions for our physical system.
If the external field and all interactions between particles are
absent, i.e., all potentials $u_l$ are constant, then the function
$h(q_1,\ldots,q_s)$ vanishes. Under this condition
$Q_N=Q_N^{(0)}=V^N$, $D_N(q_1,\ldots,q_N)=D_N^{(0)}=V^{-N}$, and
$f(q_1,\ldots,q_s)=0$. Therefore we can take $h^{(0)}=0$ and
$f^{(0)}=0$ in (\ref{F(h0,f0)})

Equation (\ref{F(h-f_2)}) and additional condition
(\ref{F(h0,f0)}) form a problem on implicit function. In
functional analysis there is a number of theorems on implicit
function for operators of various smoothness classes. We use the
theorem for analytic operator in Banach space in the form given in
the book \cite{Vainberg}

\medskip %
%\noindent %
{\bf Theorem.} (On implicit function). {\it Let ${\cal F}(h,f)$ be
an analytic operator in $D_r(h^{(0)},E_1)\times
D_{\rho}(f^{(0)},E)$ with values in $E_2$. Let an operator $B\eqd
-\partial {\cal F}(h^{(0)},f^{(0)})/\partial h$ have a bounded
inverse one. Then there are positive numbers $\rho_1$ and $r_1$
such that a unique solution $h=\chi(f)$ of the equation ${\cal
F}(h, f)=0$ with the additional condition ${\cal F}(h^{(0)},
f^{(0)})=0$ exists in a solid sphere $D_{r_1}(h^{(0)},E_1)$. This
solution is defined in a solid sphere $D_{\rho_1}(f^{(0)},E)$, is
analytic there, and satisfies the condition
$h^{(0)}=\chi(f^{(0)})$}.

\medskip %
\noindent %
Here $D_r(x_0,{\cal E})$ denotes a solid sphere of radius $r$ in a
neighborhood of the element $x_0$ in a normalized space ${\cal
E}$, the symbol $\times$ denotes the Cartesian product of sets,
$\partial {\cal F}/\partial h$ is Fr\'{e}chet derivative
\cite{Vainberg,Dunford} of the operator ${\cal F}$, $h^{(0)}$ and
$f^{(0)}$ are assigned elements of the respective spaces $E_1$ and
$E$. If the functions $h(q_1,\ldots,q_s)$ and $f(q_1,\ldots,q_s)$
satisfy this theorem conditions, the former is a single valued
operator function of the later.

To prove an existence and uniqueness of a solution of problem
(\ref{F(h-f_2)}), (\ref{F(h0,f0)}) we have to show that all
conditions of the above theorem are satisfied.

\section{Proof of feasibility of the theorem conditions \label{proof}}

First we define spaces $E_1$, $E$, and $E_2$ mentioned in the
theorem for functions describing the physical system under
consideration.

\subsection{Functional spaces of problem}

Potentials $u_k(q_1,\ldots,q_k)$ are real symmetric functions.
Suppose they are bounded below for almost all
$\{q_1,\ldots,q_k\}\in V$. All physically significant potentials
possess this property. Under this condition integral (\ref{sigma})
exists and $h(q_1,\ldots,q_s)$ is a real symmetric function
bounded for almost all $\{q_1,\ldots,q_s\}\in V$.

A set of functions bounded nearly everywhere forms a complete
linear normalized space (Banach space) with respect to the norm
\cite{Dunford,Kantorovich}
\begin{equation}
\label{norm_h} %
\|h\|=\norm{h(q_1,\ldots,q_s)},
\end{equation}                                          %  {norm_h} (17)
where "$\mathop{\rm vrai\:sup}$" denotes an essential upper bound
of the function on the indicated set. It is called the space of
essentially bounded functions and is denoted by $L_{\infty}({\cal
V}^{(s)})$. Here ${\cal V}^{(s)}\eqd V\times\cdots\times V$ is a
repeated $s$ times Cartesian product of $V$ by itself. In addition
$h(q_1,\ldots,q_s)$ satisfies condition (\ref{cond_f-h}). The set
of such functions is a subspace of $L_{\infty}({\cal V}^{(s)})$.
It is easy to show that this subspace is a complete space relative
to norm (\ref{norm_h}). Therefore we can take the Banach space of
symmetric essentially bounded functions satisfying condition
(\ref{cond_f-h}) as $E_1$.

Expression (\ref{h-f}) for $f(q_1,\ldots,q_s)$ includes multiple
integrals of different power combinations of $h(q_1,\ldots,q_s)$.
Any power of essentially bounded function are integrable with
respect to arbitrary set of variables $\{q_{j_1},\ldots,q_{j_r}\}$
over $V$ \cite{Vulikh}. Therefore all integrals in (\ref{h-f_1})
are essentially bounded functions too. Arguing as above, we can
show that the space $E$ of functions $f(q_1,\ldots,q_s)$ coincides
with $E_1$. Continuing in the same way we can show that $E_2$ is
the same space. Thus we define the spaces of the above theorem as
$E=E_1=E_2=L_{\infty}({\cal V}^{(s)})$ with property
(\ref{cond_f-h}).

From (\ref{h}) and (\ref{f}) for $h(q_1,\ldots,q_s)$ and
$f(q_1,\ldots,q_s)$ it follows that $f>-1$ and $h>-1$. Therefore
we can take a manifold \{$f>-1, \; h>-1$\} as a definition domain
of the operator ${\cal F}(h,f)$. Since the left-hand side of
(\ref{h-f_1}) is a polynomial, the operator ${\cal F}(h,f)$ is
analytical in this domain. As stated above the additional
condition (\ref{F(h0,f0)}) is valid for $h^{(0)}=f^{(0)}=0$. Thus
any solid spheres of $E_1$, $E$ with centers at $h^{(0)}=0$,
$f^{(0)}=0$ and radii $r<1$, $\rho<1$ respectively can be used as
domains $D_r(h^{(0)},E_1)$ and $D_{\rho}(f^{(0)},E)$ indicated in
the theorem.

Finally it is necessary to prove that the operator
\begin{equation}
\label{B} %
\left. B\eqd -\frac{\partial {\cal F}(h,f)}{\partial
h}\right|_{{h=0,\atop f=0\phantom{,}}\atop \vphantom{0}}
\end{equation}                                          %  {B} (18)
has a bounded inverse one.

\subsection{Properties of the operator $B$}

To find the inverse operator $B^{-1}$ it is necessary to solve the
equation $Bh=y$, where $h\in E_1$, $y\in E_2$. The expression for
$Bh$ is a linear relative to $h$ part in the left-hand side of
relation (\ref{h-f_1}) as $f=0$. Expanding products in
(\ref{h-f_1}) and keeping linear summands we obtain
\begin{equation}
\label{Bh} %
(Bh)(q_1,\ldots,q_s))=\frac{1}{V^{N-s}}\int\sum_{1\le
j_1<\cdots<j_s\le N}h(q_{j_1},\ldots,q_{j_s})dq_{s+1}\cdots dq_N.
\end{equation}                                            %  {Bh} (19)
Variables $\{q_{j_1},\ldots,q_{j_s}\}$ are divided into two groups
$\{q_{j_1},\ldots,q_{j_l}\}$ and
$\{q_{j_{\,l+1}},\ldots,q_{j_s}\}$ for $l\le s$, where
$(j_1,\ldots,j_l)\subset (1,\ldots, s)$ and
$(j_{l+1},\ldots,j_s)\subset (s+1,\ldots, N)$. The sum in the
right-hand side of (\ref{Bh}) is divided into two parts for each
$0\le l\le s$. The expression (\ref{Bh}) can be rewritten in the
form
\begin{eqnarray}
\fl (Bh)(q_1,\ldots,q_s)=\sum_{l=0}^s\;\sum_{1\le
j_1<\cdots<j_{\,l}\le s}\;\sum_{s+1\le j_{l+1}<\cdots<j_s\le N}\nonumber\\
\cdot\frac{1}{V^{N-s}}\int
h(q_{j_1},\ldots,q_{j_l},q_{j_{l+1}},\ldots,q_{j_s})dq_{s+1}\cdots
dq_N. \label{Bh1} %
\end{eqnarray}                                            %  {Bh1} (20)
We suppose that $s<N/2$. Integration in (\ref{Bh1}) is carried out
with respect to variables from second group
$\{q_{j_{\,l+1}},\ldots,q_{j_{\,s}}\}\subset \{q_{s+1},\ldots,q_N
\}$ and remaining ones $\{q_{s+1},\ldots,q_N \}\backslash
\{q_{j_{l+1}},\ldots,q_{j_s}\}$. Here a symbol $\,\backslash\,$
denotes difference of sets. Since the function $h(q_1,\ldots,q_s)$
is symmetric, we see that the summation with respect to
$(j_{l+1},\ldots,j_s)$ gives $C_{N\!-\,s}^{s-\,l}$ identical
terms. In the result we get
\begin{eqnarray}
\fl (Bh)(q_1,\ldots,q_s)=\sum_{l=1}^sC_{N\!-\,s}^{s-\,l}\;
\sum_{1\le j_1<\cdots<j_{\,l}\le s}\nonumber\\
\cdot\frac{1}{V^{s-\,l}}\int
h(q_{j_1},\ldots,q_{j_{\,l}},p_1,\ldots,p_{s-\,l})dp_1\cdots
dp_{s-\,l}. \label{Bh2} %
\end{eqnarray}                                            %  {Bh2} (21)
Here we omitted summands with $l=0$ because of property
(\ref{cond_f-h}) for the function $h$.

It's easy to estimate a norm of the operator $B$. Using definition
(\ref{norm_h}) we obtain
\begin{eqnarray}
%\begin{equation}
\fl \|Bh\|\le\sum_{l=1}^sC_{N\!-\,s}^{s-\,l}\sum_{1\le
j_1<\cdots<j_{\,l}\le
s}\|h\|\nonumber\\
=\sum_{l=1}^sC_{N\!-\,s}^{s-\,l}C_s^l\|h\|=(C_N^s-C_{N-s}^s)\|h\|.
\label{||Bh||} %
\end{eqnarray}                                            %  {||Bh||} (22)
From here we get an estimation
\begin{equation}
\label{||B||} %
\|B\|\le(C_N^s-C_{N-s}^s).
\end{equation}                                            %  {||B||} (23)
Thus the operator $B$ is bounded.

To simplify transformations we introduce notation
\begin{equation}
\label{h-(k)} %
\overline{h}^{(k)}(q_1,\ldots,q_k) = \frac{1}{V^{s-\,k}}\int
h(q_1,\ldots,q_s)dq_{k+1}\cdots dq_s.
\end{equation}                                            %  {h-(k)} (24)
Using (\ref{Bh2}) and (\ref{h-(k)}) we can write the equation
$Bh=y$ in the next form
\begin{equation}
\label{Bh3} %
\sum_{l=1}^sC_{N\!-\,s}^{s-\,l}\sum_{1\le j_1<\cdots<j_{\,l}\le
s}\;\overline{h}^{(l)}(q_{j_1},\ldots,q_{j_{\,l}})=y(q_1,\ldots,q_s).
\end{equation}                                            %  {Bh3} (25)
Integrating this equation wit respect to $q_s$, then $q_{s-1}$ and
so on we obtain a system of equations for $\overline{h}^{(k)}$
\begin{equation}
\label{Bh_k} %
\sum_{l=1}^kC_{N\!-\,k}^{s-\,l}\sum_{1\le j_1<\cdots<j_{\,l}\le
k}\;\overline{h}^{(l)}(q_{j_1},\ldots,q_{j_{\,l}})=
\overline{y}^{(k)}(q_1,\ldots,q_k)
\end{equation}                                            %  {Bh_k} (26)
for all$k=1,\ldots,s$. Correctness of this expression for
arbitrary $k$ is easily tested by induction. Note that
$\overline{h}^{(0)}=0$ and $\overline{h}^{(s)}=h$.

A solution of system (\ref{Bh_k}) is obtained step by step
starting from the first equation with $k=1$. It has the form
\begin{equation}
\overline{h}^{(k)}(q_1,\ldots,q_k)=\frac{1}{C_{N\!-\,k}^{s-\,k}}
\sum_{r=1}^k(-1)^{k-r}
\frac{C_{N\!-\,k}^{s-k+1}}{C_{N\!-\,r}^{s-k+1}}\sum_{1\le
j_1<\cdots<j_r\le k} \overline{y}^{(r)}(q_{j_1},\ldots,q_{j_r})
\label{h_l} %
\end{equation}                                            %  {h_l} (27)
for all $k=1,\ldots,s$. This solution is easily checked by forward
substitution into system (\ref{Bh_k}).

Putting here $k=s$ we obtain the expression for the inverse
operator $B^{-1}$
\begin{eqnarray}
\fl h(q_1,\ldots,q_s)=(B^{-1}y)(q_1,\ldots,q_s)\nonumber\\
=\sum_{r=1}^s(-1)^{s-r} \frac{N-s}{N-r}\sum_{1\le
j_1<\cdots<j_{\,r}\le
s}\overline{y}^{(r)}(q_{j_1},\ldots,q_{j_r}).    \label{h(y)} %
\end{eqnarray}                                            %  {h(y)} (28)
From here it's easy to estimate the norm of the inverse operator
$B^{-1}$. Evaluate the norm of right-hand-side of (\ref{h(y)})
%$$
\begin{equation}
\label{||B^-1_h||} %
\|B^{-1}y\|\le \sum_{r=1}^s \frac{N-s}{N-r}\sum_{1\le
i_1<\cdots<i_r\le s}\|y\|=\sum_{r=1}^s \frac{N-s}{N-r}C_s^r\|y\|.
\end{equation}                                       %  {||B^-1_h||} (29)
%$$
We obtain from here
\begin{equation}
\label{||B^-1||} %
\|B^{-1}\|\le \sum_{r=1}^s \frac{N-s}{N-r}C_s^r\le 2^s-1.
\end{equation}                                        %  {||B^-1||} (30)
Therefore the operator $B^{-1}$ exists and it is bounded.

So all conditions of the above theorem are valid for our physical
system. Hence there exists a unique solution $h$ of problem
(\ref{F(h-f_2)}), (\ref{F(h0,f0)}) as a function of $f$. This
solution defines an inverse transformation from the function $f$
to the function $h$: $h(q_1,\ldots,q_s)=h(q_1,\ldots,q_s;[f])$.

\section{Derivation of the inverse transformation $h(f)$ \label{sec-h(f)}}

To obtain the transformation $f\rightarrow h$ we have to solve
equation (\ref{h-f_1}) relative to $h(q_1,\ldots,q_s)$. At first
define an auxiliary operator function $g(h)$ by means of a
relation
\begin{equation}
\label{g(h)-def} %
\frac{1}{V^{N-s}}\int\prod_{1\le j_1<\cdots<j_s\le
N}[1+h(q_{j_1},\ldots,q_{j_s})]dq_{s+1}\cdots dq_N=1+g(h).
\end{equation}                                       %  {g(h)-def} (31)
This operator function is a polynomial of degree ${\cal N}=C_N^s$
relative to $h$ and depends on $s$ configurational variables
$\{q_1,\ldots,q_s\}$. It can be written in the form
\begin{equation}
\label{g(h)} %
g(h)= \sum_{l=1}^{\cal N}g_l(h),
\end{equation}                                          %  {g(h)} (32)
where $g_l(h)$ is a uniform operator of order $l$ relative to $h$.
Let us derive the expression for $g_l(h)$ from (\ref{g(h)-def}).

For the sake of abbreviation of subsequent calculations we
introduce next notations. We will denote by number
$K\in(1,\ldots,{\cal N})$ every ordered collection
$(j_1,\ldots,j_s)\subset(1,\ldots,N)$. Such one-to-one
correspondence can be always made. A collection
$(q_{j_1},\ldots,q_{j_s})$ is an element of manifold ${\cal
V}^{(s)}$. We will denote this element by $X_K$. By definition put
$X_1=(q_1,\ldots,q_s)$.

Expanding the product in (\ref{g(h)-def}) we obtain for every
$l=1,\ldots,{\cal N}$
\begin{equation}
\label{g_lh^l} %
g_l(h)=\frac{1}{V^{N-s}}\int dq_{s+1}\cdots dq_N \sum_{1\le
K_1<\cdots<K_l\le {\cal N}}h(X_{K_1})\cdots h(X_{K_l}).
\end{equation}                                         %  {g_lh^l} (33)
Introduced operators $g(h)$ and  $g_l(h)$ are symmetrical
functions of $s$ configurational variables:
$g(h)=g(q_1,\ldots,q_s;[h])$ and $g_l(h)=g_l(q_1,\ldots,q_s;[h])$.
In contrast to $h$ and $f$ both $g(q_1,\ldots,q_s;[h])$ and
$g_l(q_1,\ldots,q_s;[h])$ don't satisfy condition (\ref{cond_f-h})
except for $g_1(q_1,\ldots,q_s;[h])$. First term of series
(\ref{g(h)}) is $g_1(h)=Bh$ and satisfies to condition
(\ref{cond_f-h}). Configuration integral (\ref{Q_N1}) takes the
form
\begin{equation}
\label{Q_N2} %
Q_N=V^N[1+\overline{g}^{(0)}]=V^N[1+\sum_{k=2}^{\cal
N}\overline{g}_{k}^{(0)}].
\end{equation}                                            %  {Q_N2} (34)
Here we used notation (\ref{h-(k)}).

Substituting definitions (\ref{g(h)-def}) and (\ref{Q_N2}) into
(\ref{h-f_1}) we write it in the form
\begin{equation}
\label{F(h-f_3)} %
(1+f)[1+\overline{g}^{(0)}(h)]-[1+g(h)]=0,
\end{equation}                                            %  {F(h-f_3)} (35)
where $f$ and $g(h)$ are functions of $X_1=(q_1,\ldots,q_s)$. But
value $\overline{g}^{(0)}$ doesn't depend on configurational
variables. It is a functional relative $h$. Substituting here
expansions (\ref{g(h)}) and (\ref{Q_N2}) we reduce this equation
to the form
\begin{equation}
\label{F(h-f_4)} %
f[1+\sum_{l=2}^{\cal N}\overline{g}_l^{(0)}(h)]+\sum_{l=2}^{\cal
N}[\overline{g}_l^{(0)}(h)-g_l(h)]-Bh=0.
\end{equation}                                            %  {F(h-f_4)} (36)
Here we have taken into account that $g_1(h)=Bh$ and
$\overline{g}_1^{(0)}(h)=0$.

For subsequent calculation we need multilinear operators
\begin{equation}
\label{G_l} %
\fl G_l(y_1,\ldots,y_l)=\frac{1}{V^{N-s}}\int dq_{s+1}\cdots dq_N
\sum_{1\le K_1<\cdots<K_l\le {\cal N}}y_1(X_{K_1})\cdots
y_l(X_{K_l}).
\end{equation}                                         %  {G_l} (37)
These operators are linear with respect to any functional argument
$y_i$. We can consider the operator functions $g_l(h)$ as
generated by these multilinear operators $G_l$
\begin{equation}
\label{g_l-G_l} %
g_l(h)=G_l(h,\ldots,h).
\end{equation}                                         %  {g_l-G_l} (38)
Operators $G_l(y_1,\ldots,y_l)$ are functions of configurational
variables $(q_1,\ldots,q_s)=X_1$. In general these functions
aren't symmetrical relative to $(q_1,\ldots,q_s)$. But this isn't
important since under substituting of these operator functions
into equation (\ref{F(h-f_4)}) symmetric property will be hold
automatically. In the result we can rewrite equation
(\ref{F(h-f_4)}) as
%\begin{equation}
\begin{eqnarray}
\fl h=B^{-1}f[1+\sum_{l=2}^{\cal
N}\overline{G}_l^{(0)}(h,\ldots,h)]\nonumber\\
+B^{-1}\sum_{l=2}^{\cal N}
[\overline{G}_l^{(0)}(h,\ldots,h)-G_l(h,\ldots,h)].
\label{F(h-f_5)} %
\end{eqnarray}                                      %  {F(h-f_5)} (39)

We will search a solution of this equation in the form of power
series
\begin{equation}
\label{h(f)} %
h=\sum_{k=1}^{\infty}h_k(f),
\end{equation}                                         %  {h(f)} (40)
where $h_k(f)$ are uniform operators of order $k$ relative to $f$.
At the same time they are functions of configurational variables
$X_i$. Substituting (\ref{h(f)}) into (\ref{F(h-f_5)}) and taken
into account linearity of $G_l(y_1,\ldots,y_l)$ with respect to
any argument $y_i$ we obtain
\begin{eqnarray}
\fl \sum_{k=1}^{\infty}h_k(f)=B^{-1}f+B^{-1}\sum_{l=2}^{\cal
N}\sum_{j_1=1}^{\infty}\cdots\sum_{j_l=1}^{\infty}
\overline{G}_l^{(0)}(h_{j_1},\ldots,h_{j_l})f\nonumber\\
\label{f/G(h)1} %
+B^{-1}\sum_{l=2}^{\cal
N}\sum_{j_1=1}^{\infty}\cdots\sum_{j_l=1}^{\infty}
[\overline{G}_l^{(0)}(h_{j_1},\ldots,h_{j_l})-
G_l(h_{j_1},\ldots,h_{j_l})].
\end{eqnarray}                                      %  {f/G(h)1} (41)
Transform sums over $j_1,\ldots,j_l$ as follows
\begin{equation}
\label{sum-1}%
\sum_{j_1=1}^{\infty}\cdots\sum_{j_l=1}^{\infty}=\sum_{k=l}^{\infty}\:
\sum_{j_1+\cdots+j_l=k}.
\end{equation}                                      %  {sum-1} (42)
Then relation (\ref{f/G(h)1}) takes the form
\begin{eqnarray}
\fl \sum_{k=1}^{\infty}h_k(f)=B^{-1}f+B^{-1}\sum_{l=2}^{\cal
N}\sum_{k=l}^{\infty}\sum_{j_1+\cdots+j_{\,l}=k}
\overline{G}_l^{(0)}(h_{j_1},\ldots,h_{j_l})f\nonumber\\
\label{f/G(h)2} %
+B^{-1}\sum_{l=2}^{\cal
N}\sum_{k=l}^{\infty}\sum_{j_1+\cdots+j_l=k}
[\overline{G}_l^{(0)}(h_{j_1},\ldots,h_{j_l})-
G_l(h_{j_1},\ldots,h_{j_l})].
\end{eqnarray}                                      %  {f/G(h)2} (43)
Double sum $\sum_{l=2}^{\cal N}\sum_{k=l}^{\infty}$ is transformed
as follows%
\numparts
\begin{eqnarray}
\sum_{l=2}^{\cal N}\sum_{k=l}^{\infty}=\sum_{l=2}^{{\cal
N}-1}\sum_{k=l}^{{\cal N}-1}+\sum_{l=2}^{\cal N}\sum_{k={\cal
N}}^{\infty}=\sum_{k=2}^{{\cal N}-1}\sum_{l=2}^k+\sum_{k={\cal
N}}^{\infty}\sum_{l=2}^{\cal N} \label{sum-2}\\                %(44a)
\fl \mbox{or as follows}\nonumber\\
\sum_{l=2}^{\cal N}\sum_{k=l}^{\infty}=\sum_{l=2}^{\cal
N}\sum_{k=l}^{\cal N}+\sum_{l=2}^{\cal N}\sum_{k={\cal
N}+1}^{\infty}=\sum_{k=2}^{\cal N}\sum_{l=2}^k+\sum_{k={\cal
N}+1}^{\infty}\sum_{l=2}^{\cal N}.    \label{sum-3}         %   (44b)
\end{eqnarray}
\endnumparts
Substituting (\ref{sum-2}) and (\ref{sum-3}) into (\ref{f/G(h)2})
we get the relation
\begin{eqnarray}
\fl \sum_{k=1}^{\infty}h_k(f)=B^{-1}f+B^{-1}\sum_{k=3}^{\cal
N}\sum_{l=2}^{k-1}\sum_{j_1+\cdots+j_l=k-1}
\overline{G}_l^{(0)}(h_{j_1},\ldots,h_{j_l})f\nonumber\\
+B^{-1}\sum_{k={\cal N}+1}^{\infty}\sum_{l=2}^{\cal N}
\sum_{j_1+\cdots+j_l=k-1}
\overline{G}_l^{(0)}(h_{j_1},\ldots,h_{j_l})f\nonumber\\
+B^{-1}\sum_{k=2}^{\cal N}\sum_{l=2}^k\sum_{j_1+\cdots+j_l=k}
[\overline{G}_l^{(0)}(h_{j_1},\ldots,h_{j_l})-
G_l(h_{j_1},\ldots,h_{j_l})]\nonumber\\
+B^{-1}\sum_{k={\cal N}+1}^{\infty}\sum_{l=2}^{\cal
N}\sum_{j_1+\cdots+j_{\,l}=k}
[\overline{G}_l^{(0)}(h_{j_1},\ldots,h_{j_l})-
G_l(h_{j_1},\ldots,h_{j_l})].               \label{f/G(h)3} %
\end{eqnarray}                                      %  {f/G(h)3} (45)
Here in the first two sums of the right-hand side we change
summation variable $k$ to $k+1$.

In this relation all sums with respect to $k$ contain expressions
of order $k$ relative to $f$. Putting terms of the same order
being equal in accordance with the theorem on uniqueness of
analytical operators \cite{Hille} we obtain the next recurrent
system for the
functions $h_k(f)$ %
\numparts
\begin{eqnarray}
\label{h_1(f)}%
h_1(f)=B^{-1}f,   \\                                %  {h_1(f)} (46a)
\label{h_2(f)}%
h_2(f)=B^{-1}[\overline{G}_2^{(0)}(h_1,h_1)-G_2(h_1,h_1)],\\ % {h_2(f)} (46b)
h_k(f)=B^{-1}\sum_{l=2}^{k-1} \sum_{j_1+\cdots+j_l=k-1}
\overline{G}_l^{(0)}(h_{j_1},\ldots,h_{j_l})f+B^{-1}
\sum_{l=2}^{k}\sum_{j_1+\cdots+j_l=k}\nonumber\\  % \label{h_k(f)1}
\cdot[\overline{G}_l^{(0)}(h_{j_1},\ldots,h_{j_{\,l}})-
G_l(h_{j_1},\ldots,h_{j_l})], \qquad
3\le k\le {\cal N}, \\                                   %  {h_k(f)1} (46c)
h_k(f)=B^{-1}\sum_{l=2}^{\cal N}
\sum_{j_1+\cdots+j_l=k-1}
\overline{G}_l^{(0)}(h_{j_1},\ldots,h_{j_l})f+B^{-1}
\sum_{l=2}^{\cal N}\sum_{j_1+\cdots+j_l=k} \nonumber\\
\label{h_k(f)2}%
\cdot[\overline{G}_l^{(0)}(h_{j_1},\ldots,h_{j_{\,l}})-
G_l(h_{j_1},\ldots,h_{j_l})], \qquad k\ge{\cal N}+1.
\end{eqnarray}                                      %  {h_k(f)2} (46d)
\endnumparts %
All terms of series (\ref{h(f)}) are calculated from this system.
So the solution of the equation (\ref{F(h-f_2)}) is founded. It
satisfies additional condition (\ref{F(h0,f0)}). Convergence of
series (\ref{h(f)}) with $h_k(f)$ being the solutions of system
(\ref{h_1(f)})-(\ref{h_k(f)2}) is proved by Cauchy-Goursat method
presented in the book \cite{Vainberg}.

As soon as $h_k$ are expressed in terms of $f$ we can get the
canonical distribution (\ref{D_N1}) in terms of $F_s$ since $f$
and $F_s$ are uniquely connected by relation (\ref{f}).

\section{Calculation procedure for the canonical Gibbs distribution
 \label{sec-phi(f)}}

Since canonical distribution (\ref{D_N1}) is a ratio of two
polynomials with respect to $h$, we see that  $D_N$ is an
analytical operator function of $h$. We have just proved that $h$
is an analytical operator function of $f$. Therefore $D_N$ is an
analytical operator function of $f$ and it can be expanded into an
absolutely convergent series relative to $f$
\begin{equation}
\label{D_N(f)} %
D_N=V^{-N}\Bigl[1+ \sum_{k=1}^{\infty}\varphi_k(f)\Bigr],
\end{equation}                                       %  {D_N(f)} (47)
where $\varphi_k(f)$ is a uniform operator of order $k$
transforming function $f(q_1,\ldots,q_s)$ to function
$\varphi_k(q_1,\ldots,q_N;[f])$. Taking into account the
definitions of reduced distribution functions (\ref{F_l}) and
function $f(q_1,\ldots,q_s)$ (\ref{f}) we can get relations for
$\varphi_k$
\begin{eqnarray}
\label{phi-1-int} %
\frac{1}{V^{N-s}}\int\!dq_{s+1}\cdots dq_N
\varphi_1(q_1,\ldots,q_N;[f])=f(q_1,\ldots,q_s),\\     %  {phi-1-int} (48)
\label{phi-k-int} %
\int\!dq_{s+1}\cdots dq_N \varphi_k(q_1,\ldots,q_N;[f])=0,\qquad
k=2,3,\ldots .
\end{eqnarray}                                       %  {phi-k-int}  (49)
Below we construct a procedure for calculation of functions
$\varphi_l(q_1,\ldots,q_N;[f])$ in terms of $f$.

We introduce a nonlinear operator function $\lambda(h)$ by the
relation
\begin{equation}
\label{l(h)-def} %
\prod_{1\le j_1<\cdots<j_s\le N}
[1+h(q_{j_1},\ldots,q_{j_s})]=1+\lambda(h).
\end{equation}                                      %  {l(h)-def} (50)
This operator function is a polynomial of degree ${\cal N}$
relative to $h$. It can be written in the form
\begin{equation}
\label{l(h)} %
\lambda(h)= \sum_{k=1}^{\cal N}\lambda_k(h),
\end{equation}                                      %  {l(h)} (51)
where $\lambda_k(h)$ are defined by relations
\begin{equation}
\label{l_s(q,..)} %
\lambda_k(q_1,\ldots,q_N;[h])=\sum_{1\le K_1<\cdots<K_k\le {\cal
N}}h(X_{K_1})\cdots h(X_{K_k}).
\end{equation}                                   %  {l_s(q,..)}  (52)
 Introduce also multilinear operators
\begin{equation}
\label{L_k} %
\Lambda_k(y_1,\ldots,y_k)\eqd \sum_{1\le K_1<\cdots<K_k\le {\cal
N}}y_1(X_{K_1})\cdots y_k(X_{K_k}).
\end{equation}                                     %  {L_k}  (53)
It's evident that
\begin{equation}
\label{l_k=L_k} %
\lambda_k(h)=\Lambda_k(h,\ldots,h).
\end{equation}                                   %  {l_k=L_k}  (54)

The operators introduced here are connected with the operators
$g(h)$, $g_k(h)$, and $G_k(h_1,\ldots,h_k)$ by the relations
\begin{eqnarray}
\label{g-l} %
\frac{1}{V^{N-s}}\int dq_{s+1}\cdots
dq_N\lambda(q_1,\ldots,q_N;[h])=g(q_1,\ldots,q_s;[h]),\\  %  {g-l}  (55)
\label{g-l/k} %
\frac{1}{V^{N-s}}\int dq_{s+1}\cdots
dq_N\lambda_k(q_1,\ldots,q_N;[h])=g_k(q_1,\ldots,q_s;[h]),\\ %  {g-l/k}  (56)
\label{G-L/k} %
\frac{1}{V^{N-s}}\int dq_{s+1}\cdots
dq_N\Lambda_k(q_1,\ldots,q_N;[h_1,\ldots,h_k])\nonumber\\
=G_k(q_1,\ldots,q_s;[h_1,\ldots,h_k]).
\end{eqnarray}                                   %  {G-L/k}  (57)
In particular for $k=1$
\begin{eqnarray}
\fl \frac{1}{V^{N-s}}\int dq_{s+1}\cdots
dq_N\Lambda_1(q_1,\ldots,q_N;[h])  \nonumber\\
=G_1(q_1,\ldots,q_s;[h])=g_1(q_1,\ldots,q_s;[h])
=(Bh)(q_1,\ldots,q_s).                                    \label{G-L/1} %
\end{eqnarray}                                   %  {G-L/1}  (58)

Taking into account the expression (\ref{Q_N2}) for $Q_N$ we can
write
\begin{equation}
\label{D_N-1} %
D_N=V^{-N}\frac{1+\lambda(h)}{1+\overline{g}^{(0)}(h)}.
\end{equation}                                   %  {D_N-1}  (59)
Comparing it with (\ref{D_N(f)}) we get the relation
\begin{equation}
\label{phi(f)} %
\sum_{k=1}^{\infty}\varphi_k(f)=\frac{\lambda(h)-\overline{g}^{(0)}(h)}
{1+\overline{g}^{(0)}(h)},
\end{equation}                                   %  {phi(f)}  (60)
where $h$ is the operator function of $f$ calculated in previous
section. Using here the expressions for $\lambda(h)$,
$\overline{g}^{(0)}(h)$ and $h(f)$ we can transform the right-hand
side of (\ref{phi(f)}) to series with respect to $f$ and thus
obtain expressions for $\varphi_k(f)$. But less awkward
transformations are obtained if we construct a recurrent system
for $\varphi_k(f)$.

Multiplying (\ref{phi(f)}) by $1+\overline{g}^{(0)}(h)$ and using
(\ref{g_l-G_l}) and (\ref{l_k=L_k}) we obtain
\begin{equation}
\label{phi-G-L} %
\{1+\sum_{k=2}^{\cal
N}\overline{G}_k^{(0)}(h,...,h)\}\sum_{l=1}^{\infty}\varphi_l(f)
=\sum_{k=1}^{\cal N}\Lambda_k(h,...,h)-\sum_{k=2}^{\cal
N}\overline{G}_k^{(0)}(h,...,h).
\end{equation}                                   %  {phi-G-L}  (61)
Substitution of the expansion (\ref{h(f)}) here gives
\begin{eqnarray}
\fl \{1+\sum_{k=2}^{\cal N}\sum_{j_1=1}^{\infty}\cdots
\sum_{j_k=1}^{\infty} \overline{G}_k^{(0)}(h_{j_1},...,h_{j_k})\}
\sum_{l=1}^{\infty}\varphi_l(f)\nonumber\\
=\sum_{k=1}^{\cal N}\sum_{j_1=1}^{\infty}\cdots
\sum_{j_k=1}^{\infty}\Lambda_k(h_{j_1},...,h_{j_k})
-\sum_{k=2}^{\cal N}\sum_{j_1=1}^{\infty}\cdots
\sum_{j_k=1}^{\infty}\overline{G}_k^{(0)}(h_{j_1},...,h_{j_k}).
\label{phi-G-L_2} %
\end{eqnarray}                                   %  {phi-G-L_2}  (62)

Further calculation is carried out in the same way as in the
previous section. We won't make it and write a recurrent system
for $\varphi_k(f)$ straight away  %
\numparts
\begin{eqnarray}
\label{phi-1} %
\varphi_1(f)=\Lambda_1(h_1),\\                 %  {phi-1}  (63a)
\label{phi-2} %
\varphi_2(f)=\Lambda_1(h_2)+\Lambda_2(h_1,h_1)-
\overline{G}_2^{(0)}(h_1,h_1),\\               %  {phi-2}  (63b)
\fl \varphi_k(f)=\Lambda_1(h_k)+\sum_{l=2}^k\sum_{j_1+\cdots+j_l=k}
\{\Lambda_l(h_{j_1},...,h_{j_l})-
\overline{G}_l^{(0)}(h_{j_1},...,h_{j_l})\}\nonumber\\
\label{phi-k<N+1} %
-\sum_{l=3}^k\sum_{j_1+\cdots+j_l=k}
\overline{G}_{l-1}^{(0)}(h_{j_1},...,h_{j_{l-1}})
\varphi_{j_l}(f),\qquad 3\le
k\le {\cal N},\\                              %  {phi-k<N+1}  (63c)
\fl \varphi_k(f)=\Lambda_1(h_k)+\sum_{l=2}^{\cal
N}\sum_{j_1+\cdots+j_l=k} \{\Lambda_l(h_{j_1},...,h_{j_l})-
\overline{G}_l^{(0)}(h_{j_1},...,h_{j_l})\}\nonumber\\
\label{phi-k>N} %
-\sum_{l=3}^{{\cal N}+1}\sum_{j_1+\cdots+j_l=k}
\overline{G}_{l-1}^{(0)}(h_{j_1},...,h_{j_{l-1}})
\varphi_{j_l}(f),\qquad k\ge {\cal N}+1.
\end{eqnarray}                             %  {phi-k>N}  (63d)
\endnumparts
In this relations we have to use the expressions for $h_r(f)$
derived from the recurrent system (\ref{h_1(f)})--(\ref{h_k(f)2}).

For example an expression for $\varphi_1(f)$ is
\begin{equation}
\label{phi-1(f)} %
\varphi_1(q_1,\ldots,q_N;[f])=\sum_{r=1}^s(-1)^{s-r}
C_{N-r-1}^{s-r}\sum_{1\le j_1<\cdots<j_{\,r}\le
N}\overline{f}^{(r)}(q_{j_1},\ldots,q_{j_r}).
\end{equation}                                   %  {phi-1(f)}  (64)
It is easy to show that expressions (\ref{phi-1})--(\ref{phi-k>N})
and (\ref{phi-1(f)}) satisfy conditions (\ref{phi-1-int}),
(\ref{phi-k-int}). For s=2 expression (\ref{phi-1(f)}) coincides
with $\varphi_1(q_1,\ldots,q_N;[f])$ derived in the papers
\cite{Kalinin1,Kalinin2}.

\section{Inadequacy of reduced distribution functions of order less then
$s$ \label{inadequacy}}

Let us test the theorem conditions for reduced $l$-particle
distribution function when $l<s$. Introduce a function
$f_l(q_1,\ldots,q_l)$ by relation
\begin{equation}
\label{f_k} %
F_l(q_1,\ldots,q_l)=
\frac{N!}{(N-l)!\,V^l}[1+f_l(q_1,\ldots,q_l)].
\end{equation}                                            %  {f_k} (65)
All constructions and reasonings of sections \ref{problem} and
\ref{proof} remain valid. We obtain the operator equation ${\cal
F}_l(h,f_l)=0$ and appropriate additional condition. There are
proper Banach spaces and bounded operator $B_l$ which is
Fr\'{e}chet derivative of the nonlinear operator ${\cal
F}_l(h,f_l)$ relative to $h$. It is easy to show that the operator
$B_l$ is resulted by integrating of the operator $B$ with respect
to $q_{l+1},\ldots,q_s$. Therefore a uniform equation $B_lh=0$ has
the form (see derivation of equation (\ref{Bh3}))
\begin{equation}
\label{B_lh=0} %
\sum_{r=1}^lC_{N\!-\,l}^{s-\,r}\sum_{1\le j_1<\cdots<j_{\,r}\le
l}\;\overline{h}^{(r)}(q_{j_1},\ldots,q_{j_{\,r}})=0.
\end{equation}                                            %  {B_lh=0} (66)
From here we see that the operator $B_l$ has a nontrivial space of
zeroes. This space consists of all functions $h(q_1,\ldots,q_s)\in
E_1$ satisfying a condition
\begin{equation}
\label{h_0} %
\int h(q_1,\ldots,q_s)dq_{l+1}\cdots dq_s=0
\end{equation}                                            %  {h_0} (67)
for any fixed $l$. This means that the operator $B_l^{-1}$ isn't
exist as $l<s$ and inverse transformation $\{f_l\rightarrow h\}$
isn't exist either. Therefore we can't express the canonical
distribution $D_N$ in terms of $f_l$ as $l<s$. As soon as $l=s$
the operator $B_l=B$ and the space of zeroes of operator $B_l$
becomes trivial (see the condition (\ref{h_0})). In this case all
conditions of the theorem are valid and we obtain all above
results. So the reduced $s$-particle distribution function plays a
specific role for the system with $s$-body interaction. $F_s$ is a
reduced distribution function of minimal order containing all
information about this system.

\section{Conclusion}

Using the theorem  on implicit functions in this article it is
shown  that a reduced distribution function of order $s$ plays a
specific role for a canonical ensemble of $N$ particles with
$s$-body interaction. The canonical Gibbs distribution
$D_N(q_1\ldots,q_N)$ can be expressed uniquely in terms of this
function $F_s(q_1\ldots,q_s)$. From here we easily conclude that
there is a one-to-one correspondence between these two functions.
This means that the reduced distribution function $F_s$ contains
information about system under consideration as much as the whole
canonical distribution $D_N$. Reduced distribution functions of
all orders can be expressed in terms of this single function
$F_s$.

All reduced distribution functions of order $l$ less then $s$
don't satisfy  the theorem conditions. So it is impossible to
express the canonical distribution in terms of these functions of
order $l<s$. To all appearance they contain not all information
about the system under consideration.

Considered theorem provides sufficient conditions for existence
and uniqueness of inverse transformation $\{f\rightarrow h\}$.
Results obtained here are valid in some neighbourhood of
$h^{(0)}=0$, $f^{(0)}=0$. The question about size of this
neighbourhood demands special investigation.

\end{document}